\documentstyle[prl,aps,psfig,multicol]{revtex} 
\begin{document}
\draft
\title{The metal-insulator transition in Si:X: Anomalous response to a magnetic 
field.}
\author{M. P. Sarachik, D. Simonian, S. V. Kravchenko, and S. Bogdanovich}
\address{Physics Department, City College of the City University of 
New York, New York, New York 10031}
\author{V. Dobrosavljevic}
\address{National High Magnetic Field Laboratory, Florida State University, 
Tallahassee, Florida 32306}
\author{G. Kotliar}
\address{Serin Physics Laboratory, Rutgers University, Piscataway, New Jersey 
08854}
\date{\today}
\maketitle
\begin{abstract}
The zero-temperature magnetoconductivity of 
just-metallic Si:P scales with magnetic field, $H$, 
and dopant concentration, $n$, 
lying on a single universal curve: $\sigma(n,H)/\sigma(n,0) = 
G[H^{-\delta}\Delta n]$  with $\delta\approx 2$.    
We note that Si:P, Si:B, and Si:As all have unusually large magnetic field 
crossover exponents near 2, and suggest that this anomalously weak response to 
a magnetic field is a common feature of uncompensated
doped semiconductors.
\end{abstract}
\begin{multicols}{2}
	
The metal-insulator transition (MIT) that occurs in doped semiconductors and 
in amorphous metal-semiconductor mixtures is a continuous phase transition 
\cite{mikko,belitz,sasha}.  Some difficulty has been encountered in 
demonstrating 
the scaling that is expected to hold near such a transition.  Scaling with 
temperature and dopant concentration has been shown to hold for Si:P in the 
presence of a magnetic field of $75$  kOe \cite{belitz}.  However, the 
conductivity 
does not appear to scale with temperature in the absence of a magnetic field: 
scaling \cite{belitz2} is obtained only if one chooses a critical conductivity 
exponent, $\mu \approx 0.29$, considerably smaller than the value found  
experimentally \cite {rosenbaum,stupp,peihua}.  On the other hand, scaling of 
the zero-temperature 
conductivity has been demonstrated with magnetic field for p-type Si:B 
\cite{snezana}, 
albeit with an anomalously large magnetic field crossover exponent near 2.  
This 
raises the issue whether the anomalously weak response to a magnetic field is 
due to the spin-orbit scattering present in boron-doped silicon, or 
whether it is a general feature of uncompensated doped semiconductors near the 
metal-insulator transition.  Si:P is considered the archetypical 
strongly correlated disordered system, and is used as a standard against which 
newer materials are compared \cite{aeppli}. It is therefore of 
great fundamental interest to determine the functional form 
of the magnetoresistance close to its MIT.

To address these issues, we report measurements of the 
magnetoconductivity of Si:P.  Detailed 
analysis of data taken at low temperatures in magnetic fields to 90~kOe 
allows us to identify separate, temperature-dependent components, yielding 
reliable determinations of 
the zero-temperature conductivity.  Our results demonstrate that the 
zero-temperature conductivity scales with magnetic field and dopant 
concentrations, 
$\sigma (n,H)/\sigma(n,0) = G(H^{-\delta}\Delta n)= F (H/H^*)$, with a 
crossover exponent, $\delta\approx2$, comparable to the anomalously 
large  crossover exponent of Si:B \cite{snezana}.  Moreover, earlier data of 
Shafarman 
et al. \cite{shafarman} indicate that the magnetoconductance of Si:As strongly 
resembles 
that of Si:P, scaling with a similar crossover function and exponent.  We note 
that all the silicon-based doped semiconductors exhibit an anomalously weak 
response to a  magnetic field, and suggest that this is a feature of the 
universality class of silicon-based doped semiconductors that is currently 
not  understood.

Four Czochralski-grown Si:P samples were used in our studies with 
dopant concentrations $3.60, 3.66, 3.95$ and $4.21 \times 10^{18} 
\mbox{ cm}^{-3}$. 
Based on a critical  concentration $n_c = 3.46\times 10^{18} \mbox{ cm}^{-3}$ 
\cite{peihua}, this corresponds to 
$1.04 n_c$, $1.06 n_c$, $1.14 n_c$,  and $1.22 n_c$.  Measurements were taken 
at temperatures between 0.037~K and  0.5~K in magnetic fields up to 90~kOe.  
Sample characterization and  measurement techniques are described elsewhere 
\cite{peihua,peihua2}.

At temperatures sufficiently low that corrections due to localization are 
small, finite temperature corrections due to interactions 
are expected to yield a conductivity \cite{altsh,altsh2,lee,theory}:
\begin{equation}
\sigma(n,T)=\sigma(n,0)+A(n)T^{\frac{1}{2}}.\label{1}
\end{equation}
in the absence of a magnetic field.  
The slope $A(n)=(\sigma_{ex}-\sigma_{Har})/T^{1/2}$, where the exchange 
term, $\sigma_{ex}$, and the Hartree term, $\sigma_{Har}$, contribute with 
opposite sign.  The Hartree term consists of a $S_z=0$ channel, 
$\sigma_{s0}$, which is independent of magnetic field, and field-dependent 
$S_z=\pm1$ contributions, $\sigma_{s\pm}$, which are suppressed in very large 
fields.  Thus:
\begin{equation}
 A(n) T^{\frac{1}{2}} = (\sigma_{ex} - \sigma_{s0}) - \sigma_{s\pm}.	 
\label{2} 
\end{equation}  
For doped semiconductors in the absence of a magnetic field, the slope $A(n)$ 
is net positive near the transition and changes sign as one moves away toward 
higher dopant concentration $n$.  For the four specimens used in our study, 
the slope is  positive for the sample closest
\vbox{
\vspace{0.15in}
\hbox{
\hspace{.4in}
\psfig{file=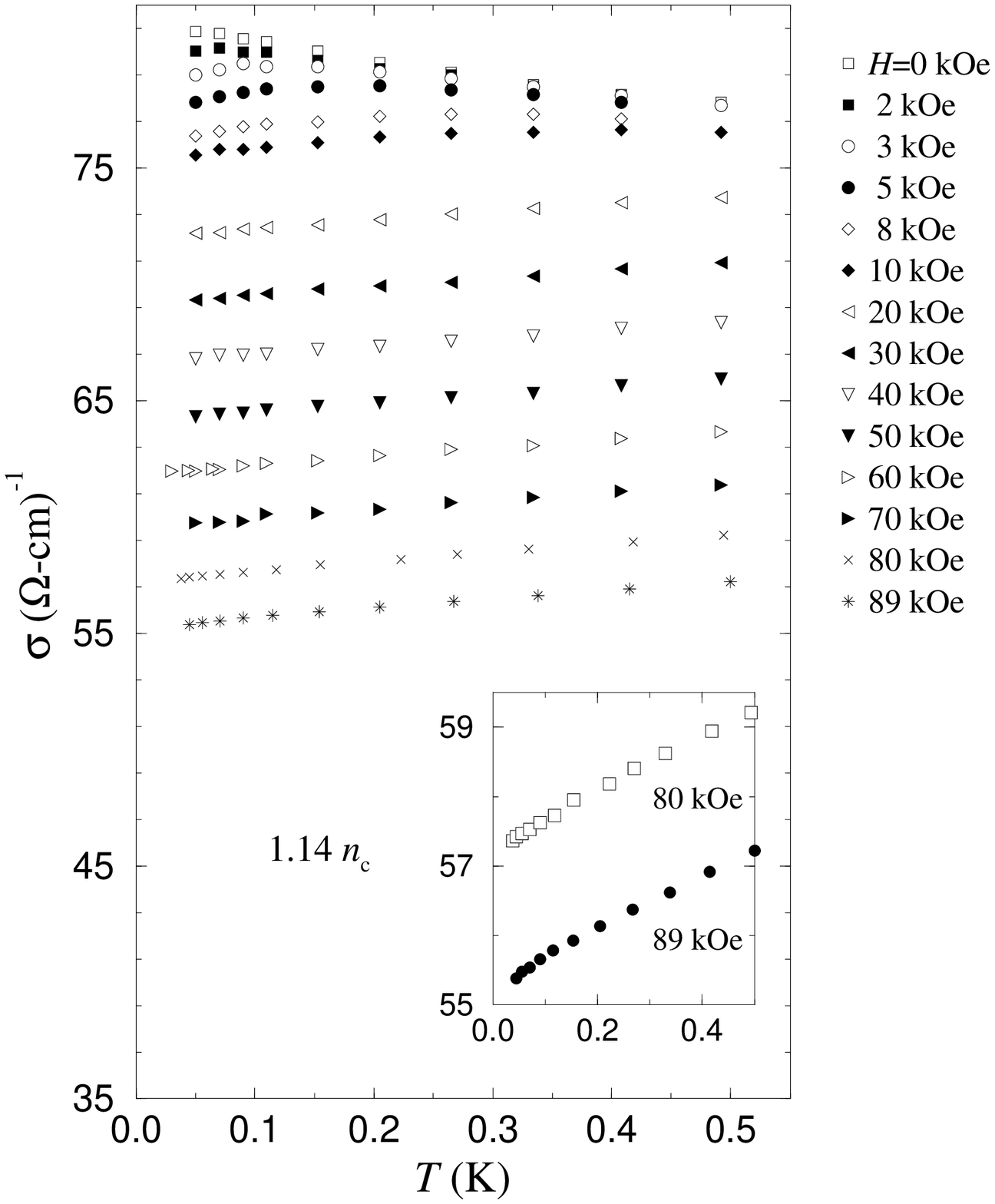,width=3.1in,bbllx=1.5in,bblly=1in,bburx=7.75in,bbury=9.25in,angle=0}
}
\vspace{0.15in}
\hbox{
\hspace{-0.15in}
\refstepcounter{figure}
\parbox[b]{3.4in}{\baselineskip=12pt \egtrm FIG.~\thefigure.
Conductivity $\sigma$ versus temperature in various magnetic fields for 
dopant concentration $n=1.14n_c$.  The inset shows data in high 
magnetic fields on an expanded scale.
\vspace{0.10in}
}
\label{fig1}
}
}
to the transition $(n=1.04n_c)$, $A\approx0$ for the sample with 
$n=1.06 n_c$, and the two samples 
furthest from the transition have net negative slopes $A$.  The samples used 
in our experiments thus span concentrations that include temperature 
coefficients for the conductivity that  are positive, zero and negative in the 
absence of a field.

The conductivity of one sample (for which A is net negative) 
is plotted in Fig.~\ref{fig1} as a function of 
temperature at various fixed magnetic fields.  The inset shows the 
conductivity in large magnetic field on an expanded scale.  The curves are 
parallel to each other in fields above 10 or 20~kOe, indicating that the 
temperature dependence in high magnetic fields is independent of the field.  
Similar behavior is found for the other three samples.  This is also 
demonstrated in Fig.~\ref{fig2}~(a), where the conductivity is shown as a 
function of 
magnetic field for one sample at four different temperatures.  Again, it is 
clear 
that the curves at different temperatures are parallel to each other in fields 
above 10 or 20~kOe.  Thus, in sufficiently high fields the 
dependence on temperature is independent of field, corresponding to the 
term $(\sigma_{ex} - \sigma_{s0}) T^{\frac{1}{2}}$  in Eq.~(2) 
above.  As shown in Fig.~\ref{fig2}~(b), all the  curves can be brought into 
coincidence in high magnetic fields by subtracting  this temperature-dependent 
term from the total conductivity.  Deviations occur in small fields, becoming 
\vbox{
\vspace{0.15in}
\hbox{
\hspace{.4in}
\psfig{file=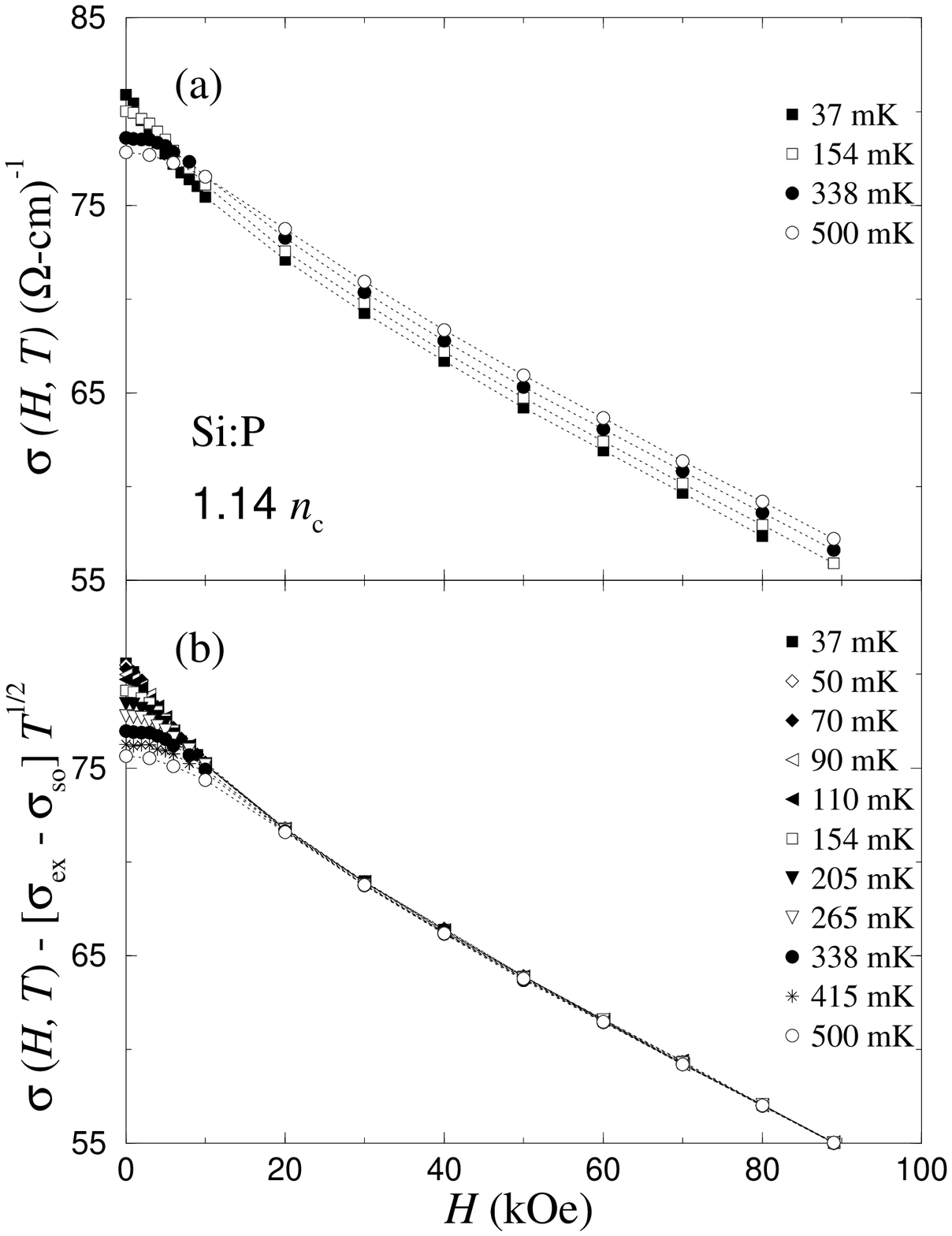,width=3.1in,bbllx=1.5in,bblly=1in,bburx=7.75in,bbury=9.25in,angle=0}
}
\vspace{0.15in}
\hbox{
\hspace{-0.15in}
\refstepcounter{figure}
\parbox[b]{3.4in}{\baselineskip=12pt \egtrm FIG.~\thefigure.
(a)  For dopant concentration $n=1.14n_c$, the conductivity $\sigma$ 
versus  magnetic field $H$ at four different temperatures, as labeled.  
(b) The conductivity minus $(\sigma_{ex} - \sigma_{s0}) T^{\frac{1}{2}}$ 
plotted as a function of magnetic field  (see text); the prefactor 
$(\sigma_{ex}-\sigma_{s0})$  varies from sample to sample.  Note  that all 
curves coincide above $10$ or $20$~kOe.
\vspace{0.10in}
}
\label{fig2}
}
}
increasingly pronounced as the temperature increases: the conductivity 
flattens below a magnetic field $H$ given approximately by $g\mu_BH\approx 
k_BT$.  This corresponds to the triplet channel contribution, $\sigma_{s\pm} 
T^{\frac{1}{2}}$, of Eq.~(2).

The curves shown in Fig.~\ref{fig2}~(b) for 37, 50 and 70~mK differ from each 
other by small amounts compared to the overall change of the 
conductivity with magnetic field; the lowest temperature data are thus close 
to the 
$T=0$ curve on this scale.  Setting $\sigma(H,0)\approx\sigma(H,37 
\text{ mK})$,  we show  $\sigma(H,0)/\sigma(0,0)$  plotted as a function of 
$H$ for all four samples in Fig.~\ref{fig3}~(a).   These four very similar 
curves can be collapsed onto a 
single curve by  appropriate choices of a scaling parameter $H^*$, as shown in 
Fig.~\ref{fig3}~(b).  We have  thereby demonstrated that the zero-temperature 
conductivity scales with  magnetic field, taking on the form 
\cite{dima,pfeuty}:
\begin{equation}
\frac{\sigma (n,H)}{\sigma (n,0)} = G (H^{-\delta}\Delta n)=
F(\frac{H}{H^*})
\label{3}
\end{equation}
with a crossover function $G (H^{-\delta}\Delta n)$, a magnetic-field 
crossover exponent  $\delta$, and a scaling parameter $H^*$ that should obey 
a power law in the critical region, $H^* \propto \Delta n^{1/\delta}$.

We note that scaling appears 
to hold quite well for the samples that have negative  as well as positive 
zero-field slopes $A$.  The crossover function of Fig.~\ref{fig3}~(b) exhibits 
complex behavior at low  fields, and becomes linear with magnetic 
field at high fields. The inset to Fig.~\ref{fig3}~(b) shows $H^*$ versus 
$\Delta n \equiv(n - n_c)$ on a log-log scale,  using $n_c =3.46 \times 10^{18} 
\mbox{ cm}^{-3}$  \cite{peihua}.  (Data are included in the inset for samples  
measured earlier and not otherwise presented in this paper.)  There appear to  
be deviations from a straight line (power law behavior) at the highest dopant 
concentration, perhaps indicating that the last sample is not in the critical  
region.  Additional careful studies are needed  to determine the breadth of 
the  critical regime.  The (inverse) slope yields a magnetic-field crossover  
exponent $\delta\approx2$, which is unusually large.
\vbox{
\vspace{0.15in}
\hbox{
\psfig{file=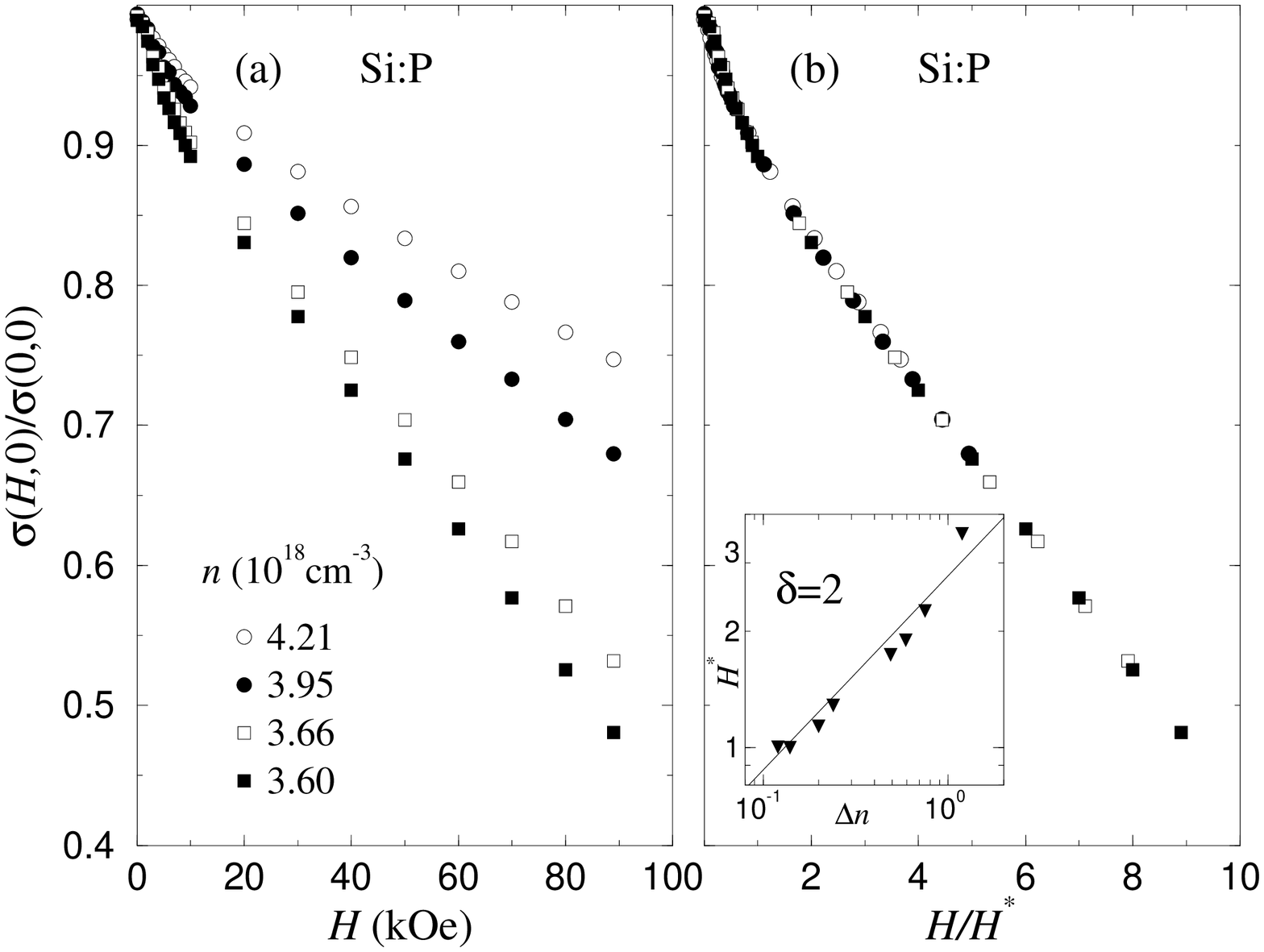,width=3.3in,bbllx=1.5in,bblly=1in,bburx=7.75in,bbury=9.25in,angle=-90}
}
\vspace{0.15in}
\hbox{
\hspace{-0.15in}
\refstepcounter{figure}
\parbox[b]{3.4in}{\baselineskip=12pt \egtrm FIG.~\thefigure.
(a)  The ratio $\sigma(H,0)/\sigma(0,0)$ versus magnetic field $H$ for 
four just-metallic Si:P samples with dopant concentrations as labeled; 
$n_c=3.46\times10^{18} \mbox{cm}^{-3}$.  (b)  Scaled curves of 
$\sigma(H,0)/\sigma(0,0)$ versus  $H/H^*$  (using $n=3.60\times10^{18} 
\mbox{cm}^{-3} = 1.04 n_c$ as the reference sample for  which $H^*=1$).  The 
inset shows $H^*$ versus  $\Delta n = (n - n_c)$ on a log-log  scale, and 
includes additional samples measured earlier.
\vspace{0.10in}
}
\label{fig3}
}
}
A similar analysis yields the curves shown for Si:B  \cite{snezana} in 
Figs.~\ref{fig4}~(a), and 
scaling is obtained for appropriate choices of the scaling parameter $H^*$, as 
shown in Fig.~\ref{fig4}~(b).  The inset is a plot of $H^*$ versus 
$\Delta n$  on a 
log-log scale.  Again, deviations from power-law behavior are apparent 
at the  higher dopant concentrations, implying that 
these are outside the 
critical range; the crossover exponent $\delta\approx1.9$.

Si:P and Si:B both have large $\delta$'s near 2.  However, their crossover 
functions are quite different, as illustrated in Fig.~\ref{fig4}~(b) 
where both are shown 
for comparison.  The effect of a magnetic field is considerably 
stronger in the case of Si:B: a 90~kOe field easily drives a just-metallic 
sample  into the insulating phase.  The magnetoconductance of Si:B exhibits 
the theoretically expected $H^{\frac{1}{2}}$ behavior \cite{lee,lee2} over a 
broad range of fields.  On the other hand, the magnetoconductance of Si:P 
displays $H^{\frac{1}{2}}$ dependence only in moderate magnetic fields and 
becomes strictly linear in $H$ at higher fields; the conductivity of Si:As 
exhibits very similar behavior \cite{shafarman} to Si:P.  We note that the 
theory of refs. \cite{lee} and \cite{lee2} is valid only  outside the critical 
region, and the behavior near the transition is not known.  The observed 
deviation from $H^{\frac{1}{2}}$ may simply reflect the fact that sufficiently 
high magnetic fields drive our samples toward the transition and into the 
critical range.  The scaling found in this paper suggests that a magnetic 
field drives samples into the critical regime for $H>H^*\propto\Delta 
n^{1/\delta}$.  

In both Si:B and Si:P, there are also clear deviations from  $H^{\frac{1}{2}}$ 
behavior in small magnetic fields: attempts to fit to this form yield a 
zero-field conductivity that is  decidedly inconsistent with the measured 
value.  This unexpected behavior in small magnetic fields is also reflected 
in anomalous temperature-dependence 
at low temperatures, and will be discussed elsewhere.

\vbox{
\vspace{0.15in}
\hbox{
\psfig{file=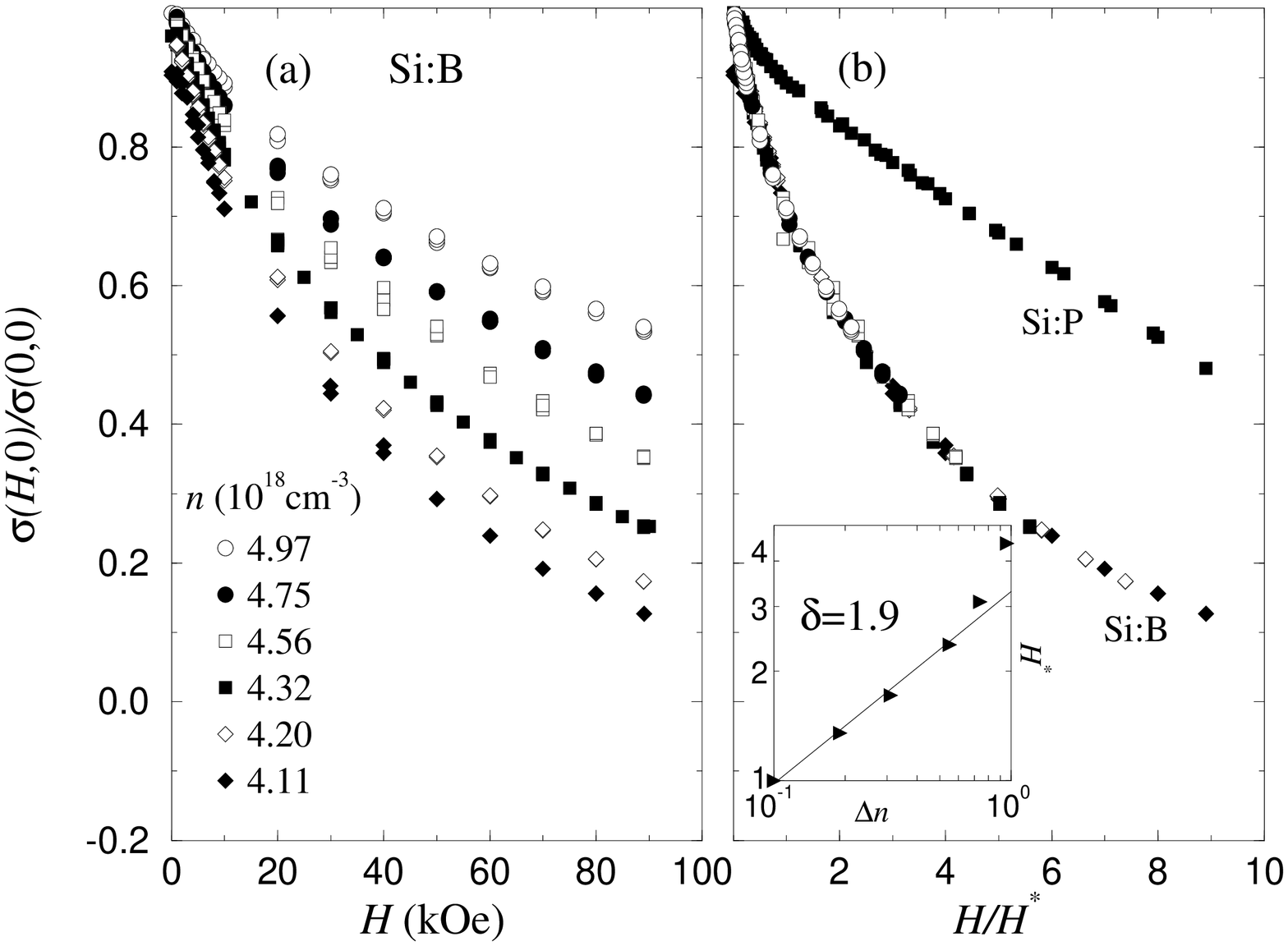,width=3.3in,bbllx=1.5in,bblly=1in,bburx=7.75in,bbury=9.25in,angle=-90}
}
\vspace{0.15in}
\hbox{
\hspace{-0.15in}
\refstepcounter{figure}
\parbox[b]{3.4in}{\baselineskip=12pt \egtrm FIG.~\thefigure.
(a)  The ratio $\sigma(H,0)/\sigma(0,0)$ versus magnetic field $H$ for
just-metallic samples of Si:B with dopant concentrations as labeled; $n_c
\approx 4.01\times 10^{18} \mbox{cm}^{-3}$.  (b) 
Scaled curves for Si:B of $\sigma(H,0)/\sigma(0,0)$ versus $H/H^*$. 
 For comparison,
the upper curve shows data for Si:P.  The inset shows $H^*$
versus $\Delta n = (n - n_c)$  on a log-log  scale (using $n=4.11\times10^{18}
\mbox{cm}^{-3} = 1.025 n_c$ as the reference sample for  which $H^*=1$).
\vspace{0.10in}
}
\label{fig4}
}
}
To summarize, the zero-temperature conductivities of Si:P and Si:B 
both scale as a function of magnetic field and dopant concentration, 
$\sigma (n,H)/\sigma (n,0) = F (H/H^*) = G (H^{-\delta}\Delta n)$, with an 
anomalously large  crossover exponent near 2.  Analysis of published data for 
Si:As \cite{shafarman} indicates that scaling is also obeyed in this system, 
again with a 
crossover exponent substantially larger than 1.  In contrast, theory predicts 
crossover exponents considerably smaller than 1: orbital effects are expected 
to give a  magnetic field crossover exponent $\delta=1/2$ \cite{dima}, and 
calculations done to date  indicate that coupling of the magnetic field 
to the electrons' spin 
yields an even smaller value  \cite{castellani,raimondi}.  Since $(\Delta n) 
\propto H^\delta$, this implies that very near 
the transition, a small  magnetic field should induce a very large 
change in critical concentration.   The large crossover exponent found in our 
experiments signals instead that for both n-type and p-type uncompensated 
doped silicon, a small magnetic field induces a small 
change in the critical concentration.  This anomalously weak response to a 
magnetic field is a feature of  the universality class of silicon-based 
doped  semiconductors that warrants further study.
  
This work was supported by the U. S. Department of Energy under 
grant no. DE-FG02-84-ER45153.

\end{multicols}

\begin{references}
\bibitem{theory} This expression is generally used to fit experimental 
results, including data for samples very close to the critical concentration.  We note, however, that its validity in the critical region has not been 
established.
\end{references}
\end{document}